\documentclass[%
    aip,
    jcp,
    amsmath,
    amssymb,
    twocolumn,longbibliography, 
    superscriptaddress,
    10pt]{revtex4-2}
\usepackage{dcolumn}
\usepackage{graphicx} 
\usepackage{bm}
\usepackage[english]{babel}
\usepackage{color}
\usepackage{amsmath} 
\usepackage{amssymb}
\usepackage{color}

\begin{document}

\title{Asymmetric Ions in Solution are Similar to Active Brownian Particles } 
\author{Setare Mostajabi Sarhangi}
\affiliation{School of Molecular Sciences, Arizona State University, PO Box 871504, Tempe, AZ 85287-1504}
\author{Dmitry V.\ Matyushov}
\email{dmitrym@asu.edu}
\affiliation{Department of Physics, School of Molecular Sciences and Center for Biological Physics, Arizona State University, PO Box 871504, Tempe, AZ 85287-1504}

\begin{abstract}
  Molecular dynamics simulations of electrolyte ions with charge shifted by the distance $d$ from the ion geometrical center have shown that mechanical equilibrium in the ion's body frame is not established beyond a critical shift distance $d^*$. A sharp crossover to a non-zero body-frame force occurs due to frustration of water molecules failing to compensate the electrostatic pull with a local density augmentation. A solution ion with asymmetric charge experiences a net body-frame force similar to self-propelled motion of active-matter Brownian particles. Thermal equilibrium and zero laboratory-frame force are still maintained, but the diffusion constant  drops significantly ($\sim 400$ times) when the charge displacement crosses the threshold value  $d^*$. Alternative routes to the diffusion constant become inequivalent and the diffusion constant from the mean-squared displacement is much higher than those from velocity and force correlation functions.         
\end{abstract}
\maketitle
\section{Introduction}
Nonpolar van der Waals (vdW) \cite{Einstein05} and electrostatic \cite{Born:1920aa} (E) interactions with the surrounding liquid are two main random forces driving an ion in solution into random motion.  The former is typically modeled in terms of the solute-solvent Lennard-Jones (LJ) interaction energy and the latter involves the interaction of the ionic charge with the fluctuating electric field of the solvent.\cite{Nee:1970} The corresponding forces, $\mathbf{F}_\text{vdW}$ and $\mathbf{F}_E$, come to mechanical equilibrium, $\mathbf{F}=\mathbf{F}_\text{vdW}+\mathbf{F}_E=0$, on the fast time scale of $\simeq 1$ fs. \cite{DMjcp1:26} The ensemble averaged forces, $\langle\tilde{\mathbf{F}}_\text{vdW}\rangle$ and $\langle\tilde{\mathbf{F}}_E\rangle$ in the body frame of the diffusing particle (denoted with tildes), can, however, be nonzero.\cite{DMjpcl:21} This comes as a result of a non-symmetric charge distribution or a non-spherical shape of the diffusing particle (solute).  The corresponding interfacial polarization, and the Maxwell electric field $\mathbf{E}$ in the interface, also become asymmetric, leading to an inhomogeneous and asymmetric local chemical potential of the solvent molecules,\cite{Landau8} $\mu = \mu_0 - c E^2$. Reaching equilibrium in response to an asymmetric $\mathbf{E}$ requires altering the field-free part of the chemical potential $\mu_0$ achieved through a local density augmentation. A compensating vdW force, produced by the density augmentation, establishes mechanical equilibrium, $\langle \tilde F^z\rangle=0$, in the body frame of the solute
\begin{equation}
	\langle \tilde F^z\rangle =  \langle \tilde F^z_E\rangle - \langle \tilde F^z_\text{vdW}\rangle . 
	\label{1}
\end{equation}             
Here, the direction of the $z$-axis in the body frame is chosen along  $\langle \tilde F^z_E\rangle$ (see below).

Body-frame mechanical equilibrium is indeed found in molecular dynamics (MD) simulations  of proteins,\cite{DMjcp:25} spherical ions, \cite{DMprl:25} and water molecules in the bulk.\cite{DMjcp1:26} However, one can anticipate that density augmentation around small ions with asymmetric charge distributions  might be limited due to local packing frustrations of solvent molecular cores. The compensation between electrostatic and density components of the chemical potential is not established in such cases, \cite{DMjpcl:21} potentially resulting in a nonzero uncompensated body-frame force in Eq.\ \eqref{1}. This Letter reports this scenario found in MD simulations of spherical LJ ions with the center of charge geometrically shifted from the center of mass, which also coincides with the point of applied LJ force. The LJ and electrostatic forces are therefore applied to different points geometrically shifted relative to each other. We find $\langle \tilde F^z\rangle\ne0$ for a sufficiently large separation between them. 

\begin{figure}
\includegraphics*[clip=true,trim= 0cm 0cm 0cm 0cm,width=9cm]{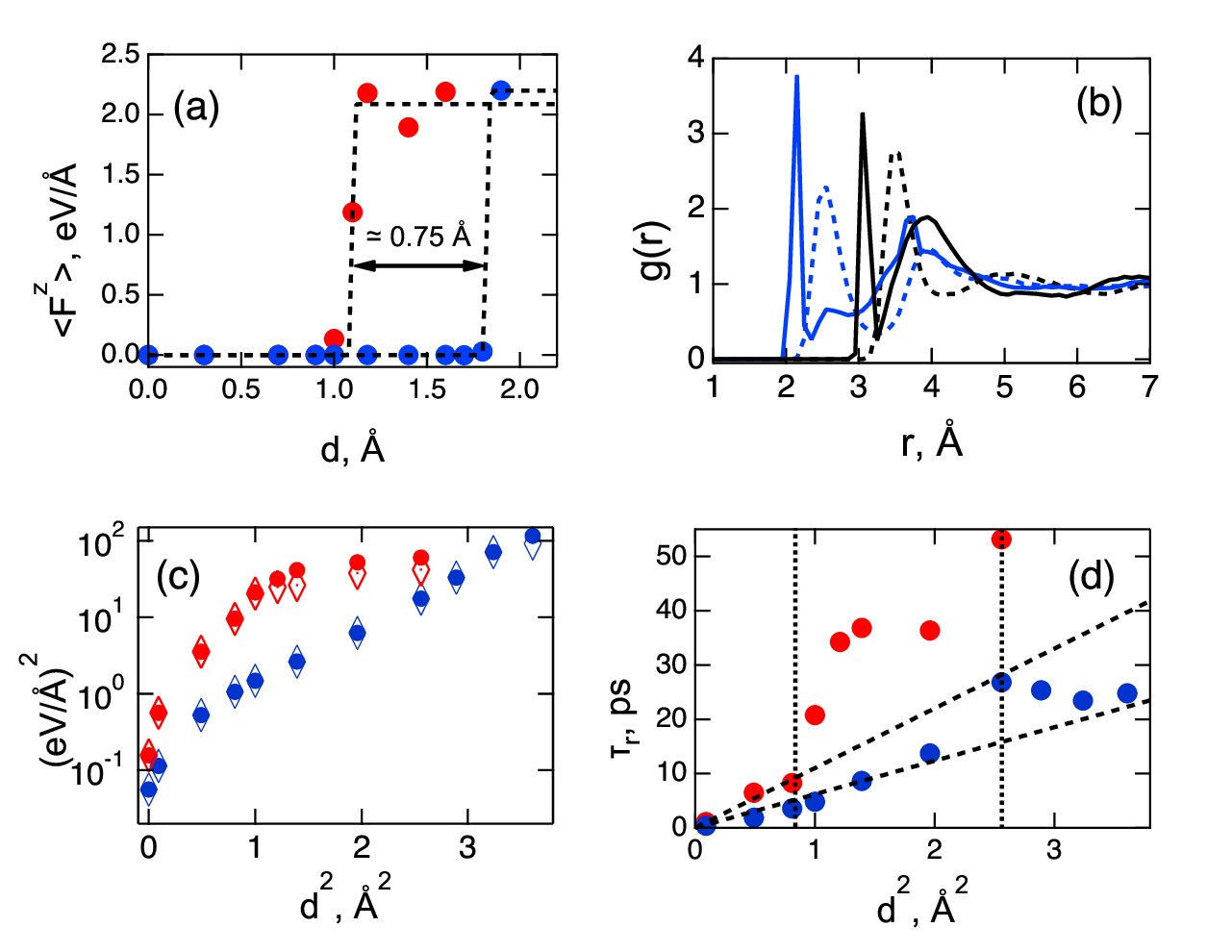}
\caption{\small (a) Average body-frame force $\langle F^z\rangle$ vs the distance $d$ of the ionic charge from the geometric center for anions (red) and cations (blue). The dashed lines are fits drawn to guide the eye. (b) Radial distribution functions $g(r)$ for the oxygen (black) and hydrogen (blue) atoms of water molecules in the  hydration shell for $d=0.3$ \AA\ (dashed lines) and $d=1.4$ \AA\ (solid lines).  (c) $\langle \mathbf{F}_E^2\rangle$ (filled points) and $-\langle \mathbf{F}_E\cdot\mathbf{F}_\text{vdW}\rangle$ (open points) vs $d^2$. (d) Rotational relaxation times $\tau_r$ of the anions (red) and cations (blue) vs $d^2$. The dashed lines are linear fits $\propto d^2$ of the initial portions of the data. The vertical dotted lines indicate the $d$ values at which SED products pass through maxima (Fig.\ \ref{fig4}d).  }
\label{fig1}
\end{figure}

The onset of a nonzero body-frame force occurs as a sharp crossover with increasing distance $d$ between the ionic charge and the solute's geometrical center (Fig.\ \ref{fig1}a). This result is found from MD simulations of cations and anions dissolved in SPC/E water. The alteration of force statistics is linked to a structural transition of water in ions' hydration shells. The hydration shell of the anion is strongly compressed beyond the crossover distance $d^*\simeq 1$ \AA. This density alteration is accompanied by the release of O-H bonds \cite{Gaffney:science2010,Davis:2012aa} pointing toward the negative ion (Fig.\ \ref{fig1}b and Supplementary Material \cite{DMZenodo2}) and is reflected in a sharp increase in variances of electrostatic and vdW forces acting on the ion (Fig.\ \ref{fig1}c). The hydration shell of the cation is more dense compared to the anion and reaching the crossover requires a stronger pull from the positive electric charge. In our simulations, we observe the crossover point for cations at $d^*\simeq 1.8$ \AA, with the phenomenology of force variances similar to that of anions.

The emergence of an uncompensated body-frame force changes dramatically the statistics and dynamics of forces acting on the diffusing particle and substantially affects ionic mobility. This force makes the equation of motion of an asymmetric ion analogous to that of an active-matter particle propelled into motion by an internal motility mechanism. Combining a constant body-frame force with rotational diffusion, the equation of motion of the off-center ion becomes that of an ``active Brownian particle''.\cite{Bechinger:2016} Its translational motion comes from taking up energy from the environment due to an imbalance (fluctuation) of microscopic forces. However, in contrast to increased mobility of self-propelled Brownian particles, the fluctuating electrostatic torque acting on the off-center charge dramatically increases dielectric friction\cite{Nee:1970,Wolynes80} on the moving ion, thus substantially lowering its diffusion constant. The energy taken from the environment is returned back, through friction, to maintain equilibrium, as is indicated by the effective temperature of the solute $T_\text{eff}$ calculated below. 

The diffusion constant turns out to be not uniquely defined for off-center ions as alternative calculation routes become inequivalent. Similarly to active Brownian particles,\cite{Bechinger:2016} the diffusion constant from the mean-squared displacement  (MSD) is much higher than corresponding diffusion constants from velocity and force correlation functions satisfying the Einstein relation.

\section{Results}
The results reported here were obtained from MD simulations of equal-size LJ cations and anions (LJ diameter of $\sigma_\text{LJ}=4.714$ \AA) in SPC/E water (Supplementary Material) with negative and positive charges shifted by the distance $d$. The direction from ion's geometrical center to the position of the charge establishes the symmetry axis associated with the $z$-axis of the body frame of reference. The direction of the $z$-axis also coincides with the direction of the average electrostatic force  $\tilde{\mathbf{F}}_E$ assigned the unit vector $\hat{\mathbf{u}}$. 

Translations of cations and anions are caused by the combined effect of LJ and electrostatic forces. On the contrary, given that the LJ potential is effectively applied to the center of mass, there is no LJ torque and LJ interactions do not affect ions' rotations. The rotational dynamics are thus driven by the fluctuating electrostatic torque $\mathbf{T}^E$ such that overdamped rotations of $\hat{\mathbf{u}}(t)$ are described by the Langevin equation, $\xi_r \dot{\hat{\mathbf{u}}} = \mathbf{T}^E\times \hat{\mathbf{u}}$, where $\xi_r$ is the rotational friction.  The average body-frame force is directed radially along the $z$-axis and thus produces no average torque. The rotational relaxation time $\tau_r$ of the solute increases continuously with its squared dipole moment\cite{DMprr2:21} $\propto d^2$ (Fig.\ \ref{fig1}d), but displays a strong upward shift at $d>d^*$. Given that the radii of the cations and anions are identical and do not change as $d$ is altered, the dependence of $\tau_r$ on the charge distribution, distinct for cations and anions, signifies the violation of standard hydrodynamic prescriptions for rotational friction (see below).  

\begin{figure}
\includegraphics*[clip=true,trim= 0cm 2cm 0cm 0cm,width=8cm]{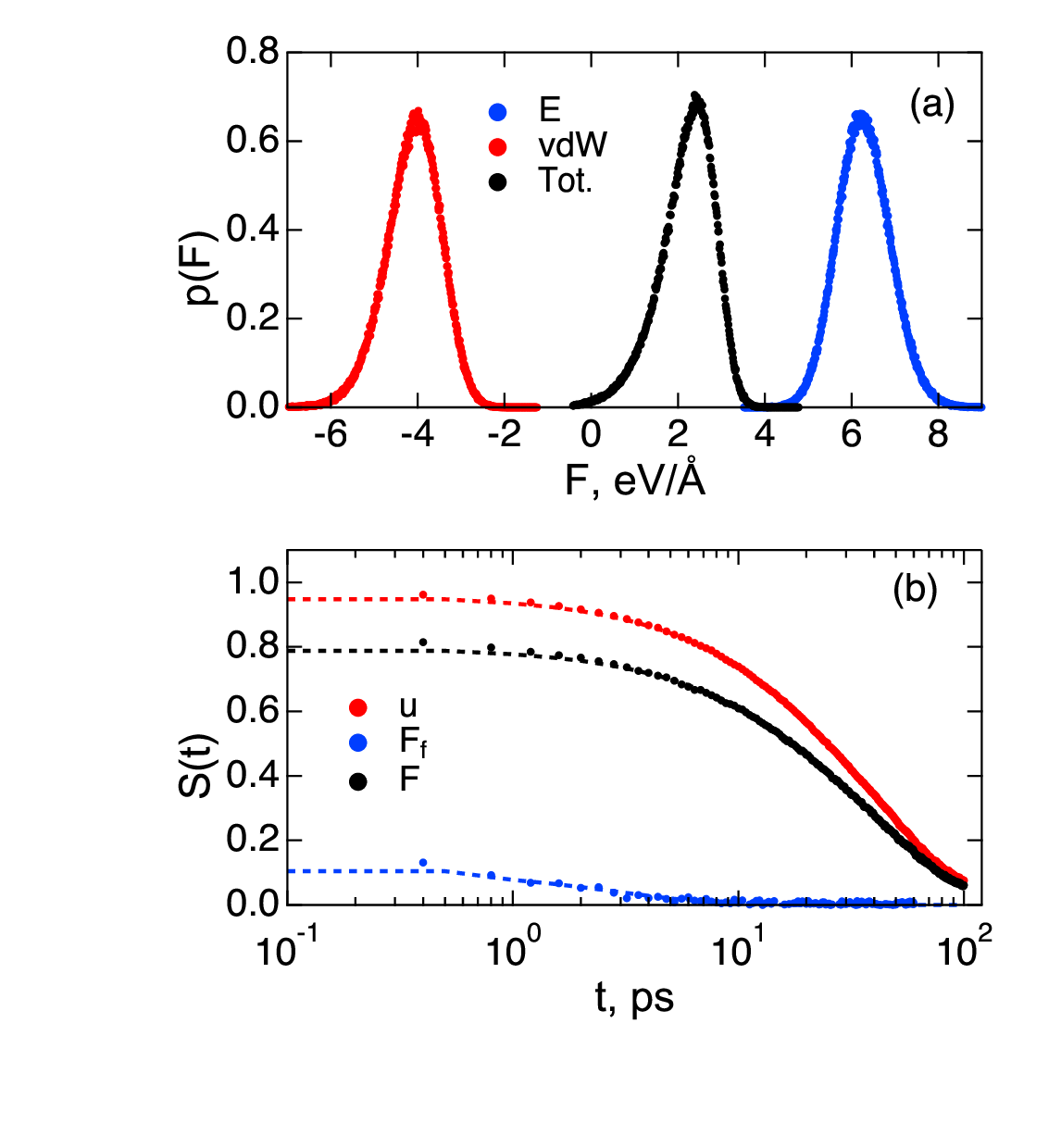}
\caption{(a) Normalized distributions of body-frame component forces $\tilde F^z_a$, $a=E,\mathrm{vdW}$ (red and blue) and of the total force $\tilde F^z$ (black) of the anion with $d=1.18$ \AA\ in SPC/E water. (b) Normalized time correlation functions of the unit vector $\hat{\mathbf{u}}(t)$ (red), of $\tilde F^z(t)$ (black), and of the fast force component $F^z_f(t)$ (blue). Dashed lines are fits to MD data (Supplementary Material). }
\label{fig2}  
\end{figure}

The distributions of electrostatic and vdW forces projected on the $z$-axis in the ion's body frame at $d>d^*$ are shown in Fig.\ \ref{fig2}a. It is clear that mechanical equilibrium is not established and there is a net uncompensated body-frame force  $\langle \tilde F^z\rangle\ne 0$ in Eq.\ \eqref{1} (black line in Fig.\ \ref{fig2}a). The total force acting on the anion carrying the off-center charge is a sum of the fast fluctuating force $\mathbf{F}_f$, $\langle \mathbf{F}_f\rangle=0$ and the average body-frame force along the direction of $\hat{\mathbf{u}}$
\begin{equation}
	\mathbf{F} = \mathbf{F}_f + \langle \tilde F^z\rangle \hat{\mathbf{u}} .
	\label{2}
\end{equation}
The presence of a non-zero average body-frame force does not imply the breakdown of the global mechanical equilibrium since the total force averages out to zero, $\langle\mathbf{F}\rangle=0$, by solute's rotations in the laboratory frame of reference, $\langle \hat{\mathbf{u}}\rangle=0$.     

The variance of the fast force component, viewed as a conservative force composed of LJ and  electrostatic components, can be related to the Laplacian of the solute-solvent LJ potential\cite{Benoir} $U_{0s}^\text{LJ}$, $\langle \mathbf{F}_f^2\rangle=k_\text{B}T\langle \nabla^2 U_{0s}^\text{LJ}\rangle$ (Supplementary Material). Assuming that the solute-solvent LJ interaction can be approximated by a harmonic potential close to the equilibrium distance, \cite{Daldrop:physrevx.7.041065} one gets for the force variance ($\langle \mathbf{F}\rangle=0$)
\begin{equation}
	\langle \mathbf{F}^2\rangle = \langle \tilde F^z\rangle^2 + 3m_ik_\text{B}T \langle \omega_\text{LJ}^2\rangle ,
	\label{3}
\end{equation}
where $m_i$ is the ionic mass and $\omega_E=\sqrt{\langle\omega_\text{LJ}^2\rangle}$ is the Einstein frequency \cite{Hansen:13} originating from the LJ cage formed by the liquid solvent around the ion: $m_i\omega_\text{LJ}^2=\partial_x^2 U_{0s}^\text{LJ}$. 

\begin{figure}
\includegraphics*[clip=true,trim= 0cm 2cm 0cm 0cm,width=8cm]{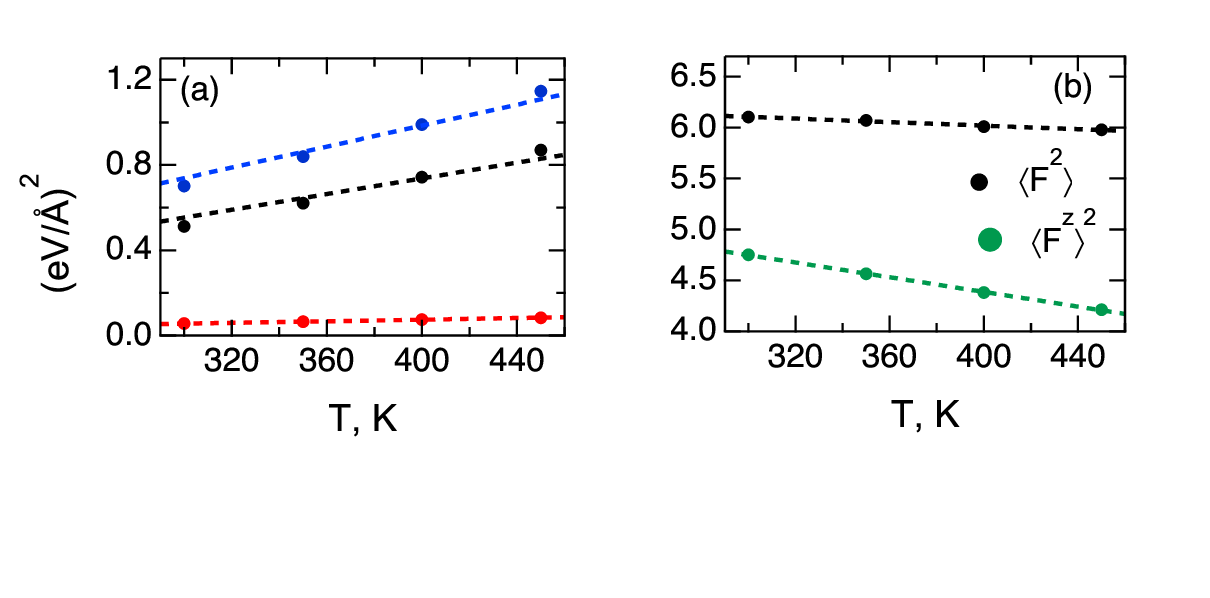}
\caption{(a) Variances of the electrostatic (red), vdW (blue), and total (black) forces for the LJ cation with $d=0$. The dashed lines are fits to linear functions with zero intercepts. (b) The total force variance (black) and $\langle \tilde F^z\rangle^2$ (green) for the LJ anion with $d=1.18$ \AA. The dashed lines are fits to linear functions.  }
\label{fig3}  
\end{figure}

Equation \eqref{3} anticipates a linear scaling of the force variance with temperature, with a nonzero intercept providing access to the uncompensated body-frame force. This prediction, with zero intercept, holds for the LJ cation at $d=0$ (Fig.\ \ref{fig3}a) resulting in the Einstein frequency $\omega_E\simeq 4.6$ ps$^{-1}$. This expectation does not hold, however, for the anion with the charge shift above the threshold value, $d>d^*$. The force variance and $\langle F^z\rangle$ decay with temperature (Fig.\ \ref{fig3}b) in that case. The corresponding $\omega_E\simeq 2.3$ ps$^{-1}$ decays with temperature as $1/T$ (Fig.\ S5 in Supplementary Material). The off-center charge of the anion hence softens the hydration shell by causing density frustration and leading to a strong temperature dependence of the effective ion-water LJ spring constant.

The force $\langle \tilde F^z\rangle\hat{\mathbf{u}}$ in Eq.\ \eqref{2} is a nonconservative force appearing as a result of statistical average, i.e., it is not represented by a gradient of the potential energy (while microscopic forces are conservative). The ability to put $\langle \tilde F^z\rangle \hat{\mathbf{u}}$ to the right-hand-side of Eq.\ \eqref{2} is based on the separation of time scales since the time of force equilibration ($\sim1$ fs) is much shorter than the corresponding times of translational and rotational diffusion.  The forces acting on the diffusing particle are conservative when $\langle \tilde F^z\rangle=0$ one arrives at a specific compensation relation between the electric force variance and the covariance of electrostatic and vdW forces 
\begin{equation}
	\langle \mathbf{F}_E^2\rangle = - \langle \mathbf{F}_E\cdot  \mathbf{F}_\text{vdW}\rangle ,
	\label{4}
\end{equation}
where $\langle \mathbf{F}_a\rangle=0$, $a=E,\mathrm{vdW}$. This relation, which is exact for electrostatic and LJ forces (Supplementary Material), was empirically established to hold with high accuracy for colloidal particles, \cite{DMjpcl2:20,Cui:2021ku} spherical ions in solution,\cite{DMprl:25} and water molecules in the bulk. \cite{Maroncelli:jocb2024,DMjcp1:26} It is equally maintained in our MD simulations for cations and anions below the threshold value of the charge displacement, $d<d^*$ (Table \ref{tab1}). These ions thus move as passive Brownian particles. In contrast, Eq.\ \eqref{4} is strongly violated when the charge is shifted to $d>d^*$ (Fig.\ \ref{fig1}c and Table \ref{tab1}). Body-frame mechanical equilibrium is not reached and $\langle \tilde F^z\rangle\ne 0$ (Tables \ref{tab1} and S6). The mobility of such ions is thus ruled by the phenomenology of active Brownian particles.

\begin{table}
\begin{ruledtabular}
\caption{{\label{tab1}} Variances of electrostatic and vdW forces ((eV/\AA)$^2$) for the total vector and $z$-projection in the body frame of the anion ($d=1.18$ \AA) and cation ($d=0$) at $T=300$ K. }
\begin{tabular}{lcccc}
Solute\footnotemark[1] & $\langle \mathbf{F}_E^2 \rangle$  & $\langle \mathbf{F}_\text{vdW}^2 \rangle$ & $\langle\mathbf{F}_E\cdot \mathbf{F}_\text{vdW} \rangle$ & $\langle \mathbf{F}^2 \rangle$\\	
\hline
Anion\footnotemark[2] & 41.2\footnotemark[3] & 17.6 & $-26.3$ & 6.1 \\
$z$-projection\footnotemark[4] & 0.38 & 0.40 & $-0.18$ & 0.42\\
Cation & 0.06  & 0.70 & $-0.06$ & 0.51 \\
\end{tabular}	
\end{ruledtabular}
\footnotetext[1]{Calculated from MD simulation of a single ion in the simulation box with $N_w=5000$ SPC/E water molecules, $d=1.18$ \AA. }
\footnotetext[2]{$\langle \tilde F^z_E\rangle=6.28$, $\langle \tilde F^z_\text{vdW}\rangle=-4.11$, and  $\langle \tilde F^z\rangle=2.18$ eV/\AA. } 
\footnotetext[3]{The value of the right-hand-side in Eq.\ \eqref{5} is 40.0 (eV/\AA)$^2$ to be compared to 41.2 (eV/\AA)$^2$ for the electric force variance. }
\footnotetext[4]{Calculated for body-frame forces $\tilde F^z_a$, $a=E,\mathrm{vdW}$ of the anion. }
\end{table}

The variances of the force components and of the total force for the anions at $d>d^*$ much exceed those for ions at $d<d^*$ (Table \ref{tab1}). This is the consequence of rotations of the uncompensated body-frame  force in the laboratory frame, e.g., $\langle \mathbf{F}^2\rangle = \langle \tilde F^z\rangle^2 + \langle \mathbf{F}_f^2\rangle$. On the contrary, when $\langle \tilde F^z\rangle=0$, the force variance, per Eq.\ \eqref{4}, becomes a difference of vdW and electrostatic self-variances: $\langle \mathbf{F}^2\rangle = \langle \mathbf{F}_\text{vdW}^2\rangle - \langle \mathbf{F}_E^2\rangle$. Electrostatic interactions thus produce effectively a negative friction, which is counter-balanced by a higher positive friction from vdW interactions. A qualitative explanation of this effect is that a spontaneous increase of the solvent density on one side of the ion increases the vdW force pushing back into the solute, but also increases the density of dipoles pulling the ion forward. The electrostatic force becomes a pulling, not friction, force. \cite{Wei:acs.jpclett.6c01311} 

Rotations of $\langle \tilde F^z\rangle$ also strongly affect the force dynamics. The overall force correlation function $\phi_F(t)\propto \langle \mathbf{F}(t)\cdot \mathbf{F}(0)\rangle$ decays on the time scale ($\tau_F\simeq 30$ ps) consistent with the dynamics of $\hat{\mathbf{u}}(t)$ ($\tau_r\simeq 37$ ps, Fig.\ \ref{fig2}b). Fluctuations of the force $\mathbf{F}_f(t)$ are much faster ($\tau_f\simeq 0.5$ ps). A similar phenomenology was found in simulations of protein diffusion:\cite{DMjpcl2:20} the dynamics of the E and vdW force components are orders of magnitude slower than of the total force because of slow rotations of average force components $\langle \tilde F^z_a\rangle$: $\langle \tilde F^z\rangle=0$ for proteins and thus only the fast force component $\mathbf{F}_f$ contributes to the dynamics of the total force. Confined environments are known to slow the rotational dynamics down \cite{Thompson:2018}  and these media thus allow separation of the component dynamics from the total force dynamics. 

The notion that fluctuations of $\mathbf{F}_f(t)$ are much faster than of the electrostatic force allows one to generalize Eq.\ \eqref{4} to the case of an uncompensated body-frame force. Since $\mathbf{F}_E(t)$ is nearly constant on the relaxation time of $\mathbf{F}_f(t)$, one can assume $\langle \mathbf{F}_f\cdot \mathbf{F}_E\rangle\simeq 0$ in Eq.\ \eqref{2}, with the result
\begin{equation}
	\langle \mathbf{F}_E^2\rangle = - \langle\mathbf{F}_E\cdot \mathbf{F}_\text{vdW}\rangle + \langle \tilde F^z\rangle \langle \hat{\mathbf{u}}\cdot \mathbf{F}_E\rangle .
	\label{5}
\end{equation}  
The generalized compensation relation in Eq.\ \eqref{5} replaces Eq.\  \eqref{4} when $\langle \tilde F^z\rangle\ne 0$. This relation is satisfied for the ions at $d>d^*$ in our simulations (Table \ref{tab1}).      

The appearance of an uncompensated body-frame force modifies a number of standard relations for the ionic mobility, making alternative routes to the ionic diffusion constant inequivalent. The generalized Langevin equation\cite{Hansen:13} provides the velocity of the solute $v_\alpha(t)$, $\alpha=x,y,z$ in terms of the memory function $M_v(t)$ and the random force $\mathbf{R}(t)$ whose initial value, $\mathbf{R}(0)=\mathbf{F}_f(0)$, is specified by the fast force component 
\begin{equation}
	\partial_t v_\alpha + \int_0^t d\tau M_v(t-\tau) v_\alpha(\tau) = m_i^{-1} R_\alpha + m_i^{-1}\langle F^z\rangle \hat{u}_\alpha   . 
	\label{15}
\end{equation}
The random force satisfies the fluctuation-dissipation relation (FDR)
\begin{equation}
	\langle \mathbf{R}(t)\cdot \mathbf{R}(0)\rangle = \langle \mathbf{F}^2\rangle m(t), 
	\label{6}
\end{equation}  
where $m(t)$ is connected to the memory function in Eq.\ \eqref{15}: $M_v(t)=\langle \mathbf{F}^2\rangle/(3m_i\langle \mathbf{v}^2\rangle)m(t)$. The function $m(t)$ is normalized for ions at $d<d^*$ ($m(0)=1$), but is not normalized at $d>d^*$ due to a substantial contribution of the body-force rotations to the force variance (Eq.\ \eqref{3}, Table \ref{tab1}). 

The Green-Kubo integral of $m(t)$ defines the memory time $\tau_m=\tilde m(0)$, where tildes over time-dependent functions are used to specify Fourier-Laplace transforms. \cite{Hansen:13} It enters the diffusion constant $D_F$ in the force route to diffusivity\cite{DMjcp:25} 
\begin{equation}
	D_F^{-1} = \tfrac{1}{3} \beta^2 \langle \mathbf{F}^2\rangle \tau_m .
	\label{7}
\end{equation}  
With the account for Eq.\ \eqref{3}, this representation specifies the friction term in Einstein's equation as $\zeta = m_i\langle \omega_\text{LJ}^2\rangle\tau_m^{-1}$ at $\langle \tilde F^z\rangle=0$. The temperature dependence of the diffusion constant is then mostly determined by $\tau_m(T)$. The advantage of the force route compared to more traditional MSD and velocity correlation routes is a direct link to forces driving diffusion in complex environments \cite{Perakis:NatCom2025} and interfaces and a significant reduction of finite-size effects in MD simulations.\cite{DMjcp:25}  

\begin{figure}
\includegraphics*[clip=true,trim= 0cm 0cm 0cm 0cm,width=8cm]{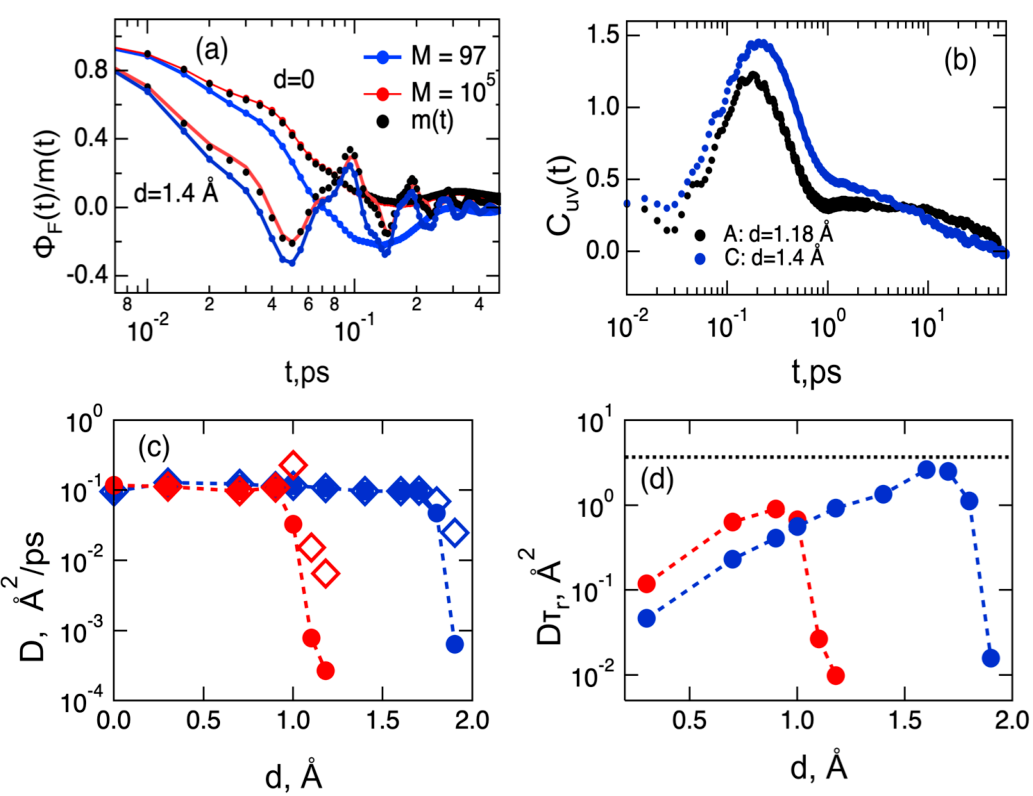} 
\caption{\small (a) Normalized force autocorrelation functions $\phi_F(t)$ for $M=97$ g/mol and $M=10^5$ g/mol cations compared to the memory function $m(t)$ calculated from numerically solving the Volterra equation (Supplementary Material). Results for $d=0$ and $d=1.4$ \AA\ are shown. (b) Time correlation function $C_{uv}(t)$ for a $M=97$ g/mol anion (A) with $d=1.18$ \AA\ and the cation (C) with $d=1.4$ \AA.  (c) Diffusion constants $D_F$ (closed points) for anions (red) and cations (blue). Open points refer to $D_\text{MSD}$ (Eq.\ \eqref{9}). (d) SED products $D\tau_r$ for cations and anions. The dotted line refers to the hydrodynamic prediction $(2/3)a_H^2$ ($a_H=\sigma_\text{LJ}/2$).   All results refer to $T=300$ K. }
\label{fig4}
\end{figure}

The memory function $m(t)$ can be calculated by solving the Volterra equation \cite{Shin:2010aa} obtained by differentiating the memory equation corresponding to Eq.\ \eqref{15}. It also follows from Eq.\ \eqref{15}  upon applying the Fourier-Laplace transform (Supplementary Material)
\begin{equation}
	\tilde m(\omega)  = \tilde{\bar{\phi}}_F(\omega)\left[1 - \frac{i\beta \langle \mathbf{F}^2\rangle }{3\omega m_i}\tilde \phi_F(\omega)\right]^{-1} ,
	\label{S16}
\end{equation}
where $\bar\phi_F(t)$ is the time cross-correlation function between the fast fluctuating force $\mathbf{F}_f(t)$ and the total force $\mathbf{F}(0)$ 
\begin{equation}
   \bar\phi_F(t) = \langle \mathbf{F}^2\rangle^{-1}\langle \mathbf{F}_f(t)\cdot \mathbf{F}(0)\rangle. 
	\label{S16}
\end{equation}
One obtains $m(t)\to \bar{\phi}_F(t)$ at $m_i\to \infty$.\cite{BalucaniBook,Shin:2010aa} This limit is used here to calculate the memory time $\tau_m$ and the diffusion constant (Eq.\  \eqref{7}) through separate MD simulations in which the masses of cations and anions were substantially increased (Supplementary Material). The Green-Kubo integral of $\bar\phi_F(t)$ then defines $\tau_m$ and $D_F$ (Table \ref{tab2}). For ions with $d<d^*$, direct calculations of the memory function based on numerical solution of the Volterra equation\cite{Shin:2010aa} are fully consistent with the force correlation function calculated for massive cations (Fig.\ \ref{fig4}a). Note that no system-size corrections were applied to the values of $D_v$ listed in Table \ref{tab2} since those tend to overestimate diffusion constant of low-mobility particles.\cite{DMjcp:25} 
  
This force route to the diffusion constant is usually equivalent to the velocity route in terms of the normalized velocity autocorrelation function $\phi_v(t)$. However, the velocity route is strongly modified by an uncompensated body-frame force. When the solute acquires a random velocity, its translational motion is accompanied by dielectric friction acting from the solvent on the off-center charge. Unless directed along the $z$-axis of the body frame, this friction force produces a torque $\mathbf{T}_E$ inducing solute rotations. The translational velocity  $\mathbf{v}(t)$ and rotation dynamics of $\hat{\mathbf{u}}(t)$ become strongly coupled as expressed through the cross-correlation function $C_{uv}(t) = \langle \hat{\mathbf{u}}(t)\cdot \mathbf{v}(0)\rangle$. This coupling modifies the solution of Eq.\ \eqref{15} for the velocity autocorrelation function and leads to the following result for the diffusion constant in the velocity route ($D_v$)
\begin{equation}
		D_v = \tfrac{1}{3} \langle \mathbf{v}^2\rangle \frac{\tilde \phi_v(0)}{1+ \tfrac{1}{3}\beta \langle \tilde F^z\rangle \tilde C_{uv}(0) }.
	\label{8}
\end{equation}

The coupling between ion's translations and rotations is the origin of a long-time tail of $C_{uv}(t)$ relaxing on the time scale of rotational motion (Fig.\ \ref{fig4}b). The Green-Kubo integral of this function, $\tilde C_{uv}(0)$ in the denominator of Eq.\ \eqref{8}, leads to a substantial reduction of $D_v$ compared to the standard result $D_v=\langle \mathbf{v}^2\rangle\tau_v/3$, $\tau_v=\tilde\phi_v(0)$. This outcome is consistent with $D_F$ (Eq.\ \eqref{7}) from MD simulations of ions with significantly enhanced masses (Table \ref{tab2}).

\begin{table}
\begin{ruledtabular}
\caption{{\label{tab2}} Diffusion constants ($\mu \mathrm{m}^2$/s) of single ions in the MD simulation box at $T=300$ K. }
\begin{tabular}{lcccccc}
Ion & $D_\text{MSD}^\text{MD}$ & $D_v$ & $D_F$ & $D_\text{MSD}$\footnotemark[1] \\	
\hline
Cation, $d=0$ & 1060\footnotemark[2] & 770\footnotemark[3]  & 960  & 960  \\
Anion, $d=1.18$ \AA  & 870 & 2.9 &  2.6 & 63    \\
\end{tabular}	
\end{ruledtabular}
\footnotetext[1]{Calculated from Eq.\ \eqref{9} with $D=D_F$. }
\footnotetext[2]{Calculated from NVT MD simulations of a single LJ cation in the simulation cell. }
\footnotetext[3]{The diffusion constant is equal to 1214 $\mu \mathrm{m}^2$/s upon adding the finite-size correction. }
\end{table}

The main result of these calculations is that ions with $d>d^*$, even though possessing masses and charges equal to ions with $d<d^*$, show a nearly $\sim 370$ drop in the diffusion constant (Table \ref{tab2}). This outcome clearly violates predictions of hydrodynamic models for molecular diffusion since neither the ion size nor the solvent viscosity has been altered when $d$ was increased. Distinctly different diffusion constants below and above $d^*$ offer a route to creating electrolytes with strong asymmetry in dynamics between cations and anions.\cite{10.1063/5.0323816} 

The Stokes-Einstein-Debye (SED) product $D\tau_r$ is predicted by hydrodynamics to be equal to $(2/3)a_H^2=3.7$ \AA$^2$ (stick boundary condition) based on the molecular ionic radius $a_H=\sigma_\text{LJ}/2$. The rotational relaxation time $\tau_r$ in the SED product is calculated as the Green-Kubo integral of the rotational time correlation function: $\phi_r^{\alpha\beta}(t)=\langle \hat{u}_\alpha(t) \hat{u}_\beta(0)\rangle = (\delta_{\alpha\beta}/3)\exp[-t/\tau_r]$. The diffusion constants for cations and anions are nearly constant with increasing $d$ below the threshold value $d^*$. The rotational relaxation time $\tau_r$ is, on the other hand, a strongly increasing function of $d$ (Fig.\ \ref{fig1}d). The SED product hence increases with $d$ until reaching the maximum at a point preceding $d^*$ (compare Figs.\ \ref{fig1}a and \ref{fig1}d), before strongly dropping due to decreased translational diffusion constant (Fig.\ \ref{fig4}d). All values of the SED product are significantly below the hydrodynamic value $(2/3)a_H^2$ (dashed horizontal line in Fig.\ \ref{fig4}d). Further, the position of SED spike is consistent with the crossover distance $d$ found for the rotational relaxation time (Fig.\ \ref{fig1}d). 
   
The long-time limit of solute's mean-squared displacement (MSD) provides an alternative route to calculate the diffusion constant. This route becomes distinct from the velocity and force routes in the presence of the body-frame force \cite{Howse:2007aa,Bechinger:2016} since solute's displacement $\Delta \mathbf{r}(t)$ becomes affected by solute's rotations. For active Brownian particles, the body-frame force leads to ``run-and-tumble'' type of trajectories\cite{Howse:2007aa} with the MSD diffusion constant enhanced relative to that from the Einstein relation. The dynamics are diffusive for both translations and rotations for asymmetric ions, but the MSD diffusion constant becomes enhanced as well. Accounting for the coupling between translations and rotations leads to the following equation (Supplementary Material)
\begin{equation}
	D_\text{MSD} = D\left[1 + \tfrac{1}{3} \beta^2\langle \tilde F^z\rangle^2D\tau_r\right] .
	\label{9}
\end{equation}   
The diffusion constant $D$ in Eq.\ \eqref{9} is either from the force (Eq.\ \eqref{7}) or from the velocity (Eq.\ \eqref{8}) routes.  Direct calculations from ionic MSDs are generally consistent with the force and velocity routes at $d<d^*$, but lead to a consistently higher $D_\text{MSD}^\text{MD}$ at $d>d^*$ (Tables \ref{tab2} and S10). The reason for this deviation is unclear to us. The results for $D_\text{MSD}$ are consistent between NVT and NVE ensembles (Supporting Material). 

``Run-and-tumble'' motions of nonequlibrium active swimmers\cite{Howse:2007aa} enhance their kinetic energy over the anticipated thermalized value. This should not happen for an ion equilibrated with the medium. One can introduce an effective temperature of ions,\cite{Bechinger:2016} $m_i\langle v_\alpha^2\rangle = k_\text{B}T_\text{eff}$. The $t\to\infty$ limit for the ion velocity follows from the corresponding Langevin equation (Eq.\ \eqref{15}) to yield the effective temperature (Supplementary Material) 
\begin{equation}
	\frac{T_\text{eff}}{T} = 1 + \frac{\beta^2 \langle \tilde F^z\rangle^2 D\tau_r}{3(1+\tau_r/(\beta Dm_i))}. 
	\label{13}
\end{equation}
Achieving $T_\text{eff}\simeq T$ for a thermalized system requires a low value of $D$ at $\langle\tilde F^z\rangle\ne 0$ in Eq.\ \eqref{13}, which is indeed found from our calculations (Tables \ref{tab2} and S10). This relation provides a consistency test, and we find no significant deviation between the kinetic temperature and $T_\text{eff}$.  

\section{Conclusions}
In conclusion, our MD simulations have demonstrated that translational diffusion can be substantially, by orders of magnitude, slowed down by geometrically separating the centers of vdW and electrostatic interactions of the diffusing particle with the medium. This slowing down, which occurs as a sharp crossover of the average body-frame force,  implies enhanced friction due to particle-medium electrostatics that goes beyond hydrodynamic models.

\acknowledgments This research was supported by the National Science Foundation (CHE-2505180). The supercomputer time was provided through allocation BIO260038 from the Advanced Cyberinfrastructure Coordination Ecosystem: Services \& Support (ACCESS) and through ASU's Research Computing.  

\section*{Data availability} The molecular-dynamics trajectory data supporting the findings of this study are available from Zenodo \cite{DMZenodo2}. The deposited dataset includes electrostatic force trajectories, Lennard-Jones force trajectories, ionic unit vector trajectories, velocity trajectories, and Python analysis scripts used to calculate the reported correlation functions. 
 
\bibliography{dielectric,dm,statmech,diffusion,glass,liquids,solvation,dynamics,simulations,surface,water,nano,ions,viscoelastisity}

\begin{thebibliography}{27}%
\makeatletter
\providecommand \@ifxundefined [1]{%
 \@ifx{#1\undefined}
}%
\providecommand \@ifnum [1]{%
 \ifnum #1\expandafter \@firstoftwo
 \else \expandafter \@secondoftwo
 \fi
}%
\providecommand \@ifx [1]{%
 \ifx #1\expandafter \@firstoftwo
 \else \expandafter \@secondoftwo
 \fi
}%
\providecommand \natexlab [1]{#1}%
\providecommand \enquote  [1]{``#1''}%
\providecommand \bibnamefont  [1]{#1}%
\providecommand \bibfnamefont [1]{#1}%
\providecommand \citenamefont [1]{#1}%
\providecommand \href@noop [0]{\@secondoftwo}%
\providecommand \href [0]{\begingroup \@sanitize@url \@href}%
\providecommand \@href[1]{\@@startlink{#1}\@@href}%
\providecommand \@@href[1]{\endgroup#1\@@endlink}%
\providecommand \@sanitize@url [0]{\catcode `\\12\catcode `\$12\catcode `\&12\catcode `\#12\catcode `\^12\catcode `\_12\catcode `\%12\relax}%
\providecommand \@@startlink[1]{}%
\providecommand \@@endlink[0]{}%
\providecommand \url  [0]{\begingroup\@sanitize@url \@url }%
\providecommand \@url [1]{\endgroup\@href {#1}{\urlprefix }}%
\providecommand \urlprefix  [0]{URL }%
\providecommand \Eprint [0]{\href }%
\providecommand \doibase [0]{https://doi.org/}%
\providecommand \selectlanguage [0]{\@gobble}%
\providecommand \bibinfo  [0]{\@secondoftwo}%
\providecommand \bibfield  [0]{\@secondoftwo}%
\providecommand \translation [1]{[#1]}%
\providecommand \BibitemOpen [0]{}%
\providecommand \bibitemStop [0]{}%
\providecommand \bibitemNoStop [0]{.\EOS\space}%
\providecommand \EOS [0]{\spacefactor3000\relax}%
\providecommand \BibitemShut  [1]{\csname bibitem#1\endcsname}%
\let\auto@bib@innerbib\@empty
\bibitem [{\citenamefont {Einstein}(2011)}]{Einstein05}%
  \BibitemOpen
  \bibfield  {author} {\bibinfo {author} {\bibfnamefont {A.}~\bibnamefont {Einstein}},\ }\href@noop {} {\emph {\bibinfo {title} {{Investigations on the Theory of the Brownian Movement}}}}\ (\bibinfo  {publisher} {BN Publishing},\ \bibinfo {year} {2011})\BibitemShut {NoStop}%
\bibitem [{\citenamefont {Born}(2 01)}]{Born:1920aa}%
  \BibitemOpen
  \bibfield  {author} {\bibinfo {author} {\bibfnamefont {M.}~\bibnamefont {Born}},\ }\bibfield  {title} {\enquote {\bibinfo {title} {{\"u}ber die beweglichkeit der elektrolytischen ionen},}\ }\href {https://doi.org/10.1007/bf01329168} {\bibfield  {journal} {\bibinfo  {journal} {Zeitschrift f{\"u}r Physik}\ }\textbf {\bibinfo {volume} {1}},\ \bibinfo {pages} {221--249} (\bibinfo {year} {1920-02-01})}\BibitemShut {NoStop}%
\bibitem [{\citenamefont {Nee}\ and\ \citenamefont {R.~Zwanzig}(1970)}]{Nee:1970}%
  \BibitemOpen
  \bibfield  {author} {\bibinfo {author} {\bibfnamefont {T.~W.}\ \bibnamefont {Nee}}\ and\ \bibinfo {author} {\bibfnamefont {R.}~\bibnamefont {R.~Zwanzig}},\ }\bibfield  {title} {\enquote {\bibinfo {title} {{Theory of dielectric relaxation in polar liquids}},}\ }\href@noop {} {\bibfield  {journal} {\bibinfo  {journal} {J. Chem. Phys.}\ }\textbf {\bibinfo {volume} {52}},\ \bibinfo {pages} {6353--6363} (\bibinfo {year} {1970})}\BibitemShut {NoStop}%
\bibitem [{\citenamefont {Pirnia}\ and\ \citenamefont {Matyushov}(2026)}]{DMjcp1:26}%
  \BibitemOpen
  \bibfield  {author} {\bibinfo {author} {\bibfnamefont {M.~M.}\ \bibnamefont {Pirnia}}\ and\ \bibinfo {author} {\bibfnamefont {D.}~\bibnamefont {Matyushov}},\ }\bibfield  {title} {\enquote {\bibinfo {title} {{Dynamics of low-temperature water are driven by electrostatics}},}\ }\href@noop {} {\bibfield  {journal} {\bibinfo  {journal} {J. Chem. Phys.}\ }\textbf {\bibinfo {volume} {164}},\ \bibinfo {pages} {144502} (\bibinfo {year} {2026})}\BibitemShut {NoStop}%
\bibitem [{\citenamefont {Samanta}, \citenamefont {Sarhangi},\ and\ \citenamefont {Matyushov}(2021)}]{DMjpcl:21}%
  \BibitemOpen
  \bibfield  {author} {\bibinfo {author} {\bibfnamefont {T.}~\bibnamefont {Samanta}}, \bibinfo {author} {\bibfnamefont {S.~M.}\ \bibnamefont {Sarhangi}},\ and\ \bibinfo {author} {\bibfnamefont {D.~V.}\ \bibnamefont {Matyushov}},\ }\bibfield  {title} {\enquote {\bibinfo {title} {{Enhanced molecular diffusivity through destructive interference between electrostatic and osmotic forces}},}\ }\href@noop {} {\bibfield  {journal} {\bibinfo  {journal} {J. Phys. Chem. Lett.}\ }\textbf {\bibinfo {volume} {12}},\ \bibinfo {pages} {6648--6653} (\bibinfo {year} {2021})}\BibitemShut {NoStop}%
\bibitem [{\citenamefont {Landau}\ and\ \citenamefont {Lifshitz}(1984)}]{Landau8}%
  \BibitemOpen
  \bibfield  {author} {\bibinfo {author} {\bibfnamefont {L.~D.}\ \bibnamefont {Landau}}\ and\ \bibinfo {author} {\bibfnamefont {E.~M.}\ \bibnamefont {Lifshitz}},\ }\href@noop {} {\emph {\bibinfo {title} {Electrodynamics of {C}ontinuous {M}edia}}}\ (\bibinfo  {publisher} {Pergamon},\ \bibinfo {address} {Oxford},\ \bibinfo {year} {1984})\BibitemShut {NoStop}%
\bibitem [{\citenamefont {Sarhangi}\ and\ \citenamefont {Matyushov}(2025{\natexlab{a}})}]{DMjcp:25}%
  \BibitemOpen
  \bibfield  {author} {\bibinfo {author} {\bibfnamefont {S.~M.}\ \bibnamefont {Sarhangi}}\ and\ \bibinfo {author} {\bibfnamefont {D.~V.}\ \bibnamefont {Matyushov}},\ }\bibfield  {title} {\enquote {\bibinfo {title} {{Memory function for protein diffusion}},}\ }\href {https://doi.org/10.1063/5.0285602} {\bibfield  {journal} {\bibinfo  {journal} {J. Chem. Phys.}\ }\textbf {\bibinfo {volume} {163}},\ \bibinfo {pages} {095101} (\bibinfo {year} {2025}{\natexlab{a}})}\BibitemShut {NoStop}%
\bibitem [{\citenamefont {Huang}\ \emph {et~al.}(2025)\citenamefont {Huang}, \citenamefont {Sarhangi}, \citenamefont {Granick},\ and\ \citenamefont {Matyushov}}]{DMprl:25}%
  \BibitemOpen
  \bibfield  {author} {\bibinfo {author} {\bibfnamefont {T.}~\bibnamefont {Huang}}, \bibinfo {author} {\bibfnamefont {S.~M.}\ \bibnamefont {Sarhangi}}, \bibinfo {author} {\bibfnamefont {S.}~\bibnamefont {Granick}},\ and\ \bibinfo {author} {\bibfnamefont {D.~V.}\ \bibnamefont {Matyushov}},\ }\bibfield  {title} {\enquote {\bibinfo {title} {{Aqueous ion mobility over a broad concentration range}},}\ }\href@noop {} {\bibfield  {journal} {\bibinfo  {journal} {Phys. Rev. Lett.}\ }\textbf {\bibinfo {volume} {135}},\ \bibinfo {pages} {028002} (\bibinfo {year} {2025})}\BibitemShut {NoStop}%
\bibitem [{\citenamefont {Ji}, \citenamefont {Odelius},\ and\ \citenamefont {Gaffney}(2010)}]{Gaffney:science2010}%
  \BibitemOpen
  \bibfield  {author} {\bibinfo {author} {\bibfnamefont {M.}~\bibnamefont {Ji}}, \bibinfo {author} {\bibfnamefont {M.}~\bibnamefont {Odelius}},\ and\ \bibinfo {author} {\bibfnamefont {K.~J.}\ \bibnamefont {Gaffney}},\ }\bibfield  {title} {\enquote {\bibinfo {title} {{Large angular jump mechanism observed for hydrogen bond exchange in aqueous perchlorate solution}},}\ }\href {https://doi.org/10.1126/science.1187707} {\bibfield  {journal} {\bibinfo  {journal} {Science}\ }\textbf {\bibinfo {volume} {328}},\ \bibinfo {pages} {1003--1005} (\bibinfo {year} {2010})}\BibitemShut {NoStop}%
\bibitem [{\citenamefont {Davis}\ \emph {et~al.}(1 21)\citenamefont {Davis}, \citenamefont {Gierszal}, \citenamefont {Wang},\ and\ \citenamefont {Ben-Amotz}}]{Davis:2012aa}%
  \BibitemOpen
  \bibfield  {author} {\bibinfo {author} {\bibfnamefont {J.~G.}\ \bibnamefont {Davis}}, \bibinfo {author} {\bibfnamefont {K.~P.}\ \bibnamefont {Gierszal}}, \bibinfo {author} {\bibfnamefont {P.}~\bibnamefont {Wang}},\ and\ \bibinfo {author} {\bibfnamefont {D.}~\bibnamefont {Ben-Amotz}},\ }\bibfield  {title} {\enquote {\bibinfo {title} {Water structural transformation at molecular hydrophobic interfaces},}\ }\href {https://doi.org/10.1038/nature11570} {\bibfield  {journal} {\bibinfo  {journal} {Nature}\ }\textbf {\bibinfo {volume} {491}},\ \bibinfo {pages} {582 -- 585} (\bibinfo {year} {2012-11-21})}\BibitemShut {NoStop}%
\bibitem [{\citenamefont {Sarhangi}\ and\ \citenamefont {Matyushov}(2025{\natexlab{b}})}]{DMZenodo2}%
  \BibitemOpen
  \bibfield  {author} {\bibinfo {author} {\bibfnamefont {S.~M.}\ \bibnamefont {Sarhangi}}\ and\ \bibinfo {author} {\bibfnamefont {D.~V.}\ \bibnamefont {Matyushov}},\ }\href@noop {} {\enquote {\bibinfo {title} {10.5281/zenodo.20451478},}\ } (\bibinfo {year} {2025}{\natexlab{b}})\BibitemShut {NoStop}%
\bibitem [{\citenamefont {Bechinger}\ \emph {et~al.}(2016)\citenamefont {Bechinger}, \citenamefont {Leonardo}, \citenamefont {L{\"o}wen}, \citenamefont {Reichhardt}, \citenamefont {Volpe},\ and\ \citenamefont {Volpe}}]{Bechinger:2016}%
  \BibitemOpen
  \bibfield  {author} {\bibinfo {author} {\bibfnamefont {C.}~\bibnamefont {Bechinger}}, \bibinfo {author} {\bibfnamefont {R.~D.}\ \bibnamefont {Leonardo}}, \bibinfo {author} {\bibfnamefont {H.}~\bibnamefont {L{\"o}wen}}, \bibinfo {author} {\bibfnamefont {C.}~\bibnamefont {Reichhardt}}, \bibinfo {author} {\bibfnamefont {G.}~\bibnamefont {Volpe}},\ and\ \bibinfo {author} {\bibfnamefont {G.}~\bibnamefont {Volpe}},\ }\bibfield  {title} {\enquote {\bibinfo {title} {{Active particles in complex and crowded environments}},}\ }\href {https://doi.org/10.1103/revmodphys.88.045006} {\bibfield  {journal} {\bibinfo  {journal} {Rev. Mod. Phys.}\ }\textbf {\bibinfo {volume} {88}},\ \bibinfo {pages} {045006} (\bibinfo {year} {2016})}\BibitemShut {NoStop}%
\bibitem [{\citenamefont {Wolynes}(1980)}]{Wolynes80}%
  \BibitemOpen
  \bibfield  {author} {\bibinfo {author} {\bibfnamefont {P.}~\bibnamefont {Wolynes}},\ }\bibfield  {title} {\enquote {\bibinfo {title} {{Dynamics of electrolyte solutions}},}\ }\href@noop {} {\bibfield  {journal} {\bibinfo  {journal} {Ann. Rev. Phys. Chem.}\ }\textbf {\bibinfo {volume} {31}},\ \bibinfo {pages} {345--376} (\bibinfo {year} {1980})}\BibitemShut {NoStop}%
\bibitem [{\citenamefont {Samanta}\ and\ \citenamefont {Matyushov}(2021)}]{DMprr2:21}%
  \BibitemOpen
  \bibfield  {author} {\bibinfo {author} {\bibfnamefont {T.}~\bibnamefont {Samanta}}\ and\ \bibinfo {author} {\bibfnamefont {D.~V.}\ \bibnamefont {Matyushov}},\ }\bibfield  {title} {\enquote {\bibinfo {title} {{Dielectric friction, violation of the Stokes-Einstein-Debye relation, and non-Gaussian transport dynamics of dipolar solutes in water}},}\ }\href@noop {} {\bibfield  {journal} {\bibinfo  {journal} {Phys. Rev. Res.}\ }\textbf {\bibinfo {volume} {3}},\ \bibinfo {pages} {023025} (\bibinfo {year} {2021})}\BibitemShut {NoStop}%
\bibitem [{Ben()}]{Benoir}%
  \BibitemOpen
  \href@noop {} {\enquote {\bibinfo {title} {{The connection between the force variance and the second derivative of LJ potential was communicated to us by Benoit Roux}},}\ }\BibitemShut {NoStop}%
\bibitem [{\citenamefont {Daldrop}, \citenamefont {Kowalik},\ and\ \citenamefont {Netz}(2017)}]{Daldrop:physrevx.7.041065}%
  \BibitemOpen
  \bibfield  {author} {\bibinfo {author} {\bibfnamefont {J.~O.}\ \bibnamefont {Daldrop}}, \bibinfo {author} {\bibfnamefont {B.~G.}\ \bibnamefont {Kowalik}},\ and\ \bibinfo {author} {\bibfnamefont {R.~R.}\ \bibnamefont {Netz}},\ }\bibfield  {title} {\enquote {\bibinfo {title} {{External potential modifies friction of molecular solutes in water}},}\ }\href {https://doi.org/10.1103/physrevx.7.041065} {\bibfield  {journal} {\bibinfo  {journal} {Phys. Rev. X}\ }\textbf {\bibinfo {volume} {7}},\ \bibinfo {pages} {041065} (\bibinfo {year} {2017})}\BibitemShut {NoStop}%
\bibitem [{\citenamefont {Hansen}\ and\ \citenamefont {McDonald}(2013)}]{Hansen:13}%
  \BibitemOpen
  \bibfield  {author} {\bibinfo {author} {\bibfnamefont {J.-P.}\ \bibnamefont {Hansen}}\ and\ \bibinfo {author} {\bibfnamefont {I.~R.}\ \bibnamefont {McDonald}},\ }\href@noop {} {\emph {\bibinfo {title} {Theory of {S}imple {L}iquids}}},\ \bibinfo {edition} {4th}\ ed.\ (\bibinfo  {publisher} {Academic Press},\ \bibinfo {address} {Amsterdam},\ \bibinfo {year} {2013})\BibitemShut {NoStop}%
\bibitem [{\citenamefont {Sarhangi}\ and\ \citenamefont {Matyushov}(2020)}]{DMjpcl2:20}%
  \BibitemOpen
  \bibfield  {author} {\bibinfo {author} {\bibfnamefont {S.~M.}\ \bibnamefont {Sarhangi}}\ and\ \bibinfo {author} {\bibfnamefont {D.~V.}\ \bibnamefont {Matyushov}},\ }\bibfield  {title} {\enquote {\bibinfo {title} {{Driving forces of protein diffusivity}},}\ }\href@noop {} {\bibfield  {journal} {\bibinfo  {journal} {J. Phys. Chem. Lett.}\ }\textbf {\bibinfo {volume} {11}},\ \bibinfo {pages} {10137--10143} (\bibinfo {year} {2020})}\BibitemShut {NoStop}%
\bibitem [{\citenamefont {Cui}\ and\ \citenamefont {Cui}(2021)}]{Cui:2021ku}%
  \BibitemOpen
  \bibfield  {author} {\bibinfo {author} {\bibfnamefont {A.~Y.}\ \bibnamefont {Cui}}\ and\ \bibinfo {author} {\bibfnamefont {Q.}~\bibnamefont {Cui}},\ }\bibfield  {title} {\enquote {\bibinfo {title} {{Modulation of nanoparticle diffusion by surface ligand length and charge: Analysis with molecular dynamics simulations}},}\ }\href@noop {} {\bibfield  {journal} {\bibinfo  {journal} {J. Phys. Chem. B}\ }\textbf {\bibinfo {volume} {125}},\ \bibinfo {pages} {4555--4565} (\bibinfo {year} {2021})}\BibitemShut {NoStop}%
\bibitem [{\citenamefont {Mukherjee}, \citenamefont {Palchowdhury},\ and\ \citenamefont {Maroncelli}(2024)}]{Maroncelli:jocb2024}%
  \BibitemOpen
  \bibfield  {author} {\bibinfo {author} {\bibfnamefont {K.}~\bibnamefont {Mukherjee}}, \bibinfo {author} {\bibfnamefont {S.}~\bibnamefont {Palchowdhury}},\ and\ \bibinfo {author} {\bibfnamefont {M.}~\bibnamefont {Maroncelli}},\ }\bibfield  {title} {\enquote {\bibinfo {title} {{Do electrostatics control the diffusive dynamics of solitary water? NMR and MD studies of water translation and rotation in dipolar and ionic solvents}},}\ }\href {https://doi.org/10.1021/acs.jpcb.3c08300} {\bibfield  {journal} {\bibinfo  {journal} {J. Phys. Chem. B}\ }\textbf {\bibinfo {volume} {128}},\ \bibinfo {pages} {3689--3706} (\bibinfo {year} {2024})}\BibitemShut {NoStop}%
\bibitem [{\citenamefont {Wei}\ \emph {et~al.}(2026)\citenamefont {Wei}, \citenamefont {Chen}, \citenamefont {Ren}, \citenamefont {He}, \citenamefont {Xu}, \citenamefont {Liu}, \citenamefont {Zheng}, \citenamefont {Zhang}, \citenamefont {Si}, \citenamefont {Sha}, \citenamefont {Ni},\ and\ \citenamefont {Chen}}]{Wei:acs.jpclett.6c01311}%
  \BibitemOpen
  \bibfield  {author} {\bibinfo {author} {\bibfnamefont {Z.}~\bibnamefont {Wei}}, \bibinfo {author} {\bibfnamefont {M.}~\bibnamefont {Chen}}, \bibinfo {author} {\bibfnamefont {J.}~\bibnamefont {Ren}}, \bibinfo {author} {\bibfnamefont {P.}~\bibnamefont {He}}, \bibinfo {author} {\bibfnamefont {W.}~\bibnamefont {Xu}}, \bibinfo {author} {\bibfnamefont {W.}~\bibnamefont {Liu}}, \bibinfo {author} {\bibfnamefont {F.}~\bibnamefont {Zheng}}, \bibinfo {author} {\bibfnamefont {Y.}~\bibnamefont {Zhang}}, \bibinfo {author} {\bibfnamefont {W.}~\bibnamefont {Si}}, \bibinfo {author} {\bibfnamefont {J.}~\bibnamefont {Sha}}, \bibinfo {author} {\bibfnamefont {Z.}~\bibnamefont {Ni}},\ and\ \bibinfo {author} {\bibfnamefont {Y.}~\bibnamefont {Chen}},\ }\bibfield  {title} {\enquote {\bibinfo {title} {{A microscopic origin for the breakdown of the Stokes--Einstein relation in ion transport}},}\ }\href {https://doi.org/10.1021/acs.jpclett.6c01311} {\bibfield  {journal} {\bibinfo  {journal} {J. Phys. Chem. Lett.}\ }\textbf {\bibinfo {volume} {17}},\ \bibinfo {pages} {6662--6667} (\bibinfo {year} {2026})}\BibitemShut {NoStop}%
\bibitem [{\citenamefont {Thompson}(2018)}]{Thompson:2018}%
  \BibitemOpen
  \bibfield  {author} {\bibinfo {author} {\bibfnamefont {W.~H.}\ \bibnamefont {Thompson}},\ }\bibfield  {title} {\enquote {\bibinfo {title} {{Perspective: Dynamics of confined liquids}},}\ }\href {https://doi.org/10.1063/1.5057759} {\bibfield  {journal} {\bibinfo  {journal} {J. Chem. Phys.}\ }\textbf {\bibinfo {volume} {149}},\ \bibinfo {pages} {170901} (\bibinfo {year} {2018})}\BibitemShut {NoStop}%
\bibitem [{\citenamefont {Girelli}\ \emph {et~al.}(2025)\citenamefont {Girelli}, \citenamefont {Bin}, \citenamefont {Filianina}, \citenamefont {Dargasz}, \citenamefont {Anthuparambil}, \citenamefont {M{\"o}ller}, \citenamefont {Zozulya}, \citenamefont {Andronis}, \citenamefont {Timmermann}, \citenamefont {Berkowicz}, \citenamefont {Retzbach}, \citenamefont {Reiser}, \citenamefont {Raza}, \citenamefont {Kowalski}, \citenamefont {Akhundzadeh}, \citenamefont {Schrage}, \citenamefont {Woo}, \citenamefont {Senft}, \citenamefont {Reichart}, \citenamefont {Leonau}, \citenamefont {Prince}, \citenamefont {Ch{\`e}vremont}, \citenamefont {Seydel}, \citenamefont {Hallmann}, \citenamefont {Rodriguez-Fernandez}, \citenamefont {Pudell}, \citenamefont {Brausse}, \citenamefont {Boesenberg}, \citenamefont {Wrigley}, \citenamefont {Youssef}, \citenamefont {Lu}, \citenamefont {Jo}, \citenamefont {Shayduk}, \citenamefont {Guest}, \citenamefont {Madsen}, \citenamefont {Lehmk{\"u}hler}, \citenamefont {Paulus}, \citenamefont {Zhang}, \citenamefont {Schreiber}, \citenamefont {Gutt},\ and\ \citenamefont {Perakis}}]{Perakis:NatCom2025}%
  \BibitemOpen
  \bibfield  {author} {\bibinfo {author} {\bibfnamefont {A.}~\bibnamefont {Girelli}}, \bibinfo {author} {\bibfnamefont {M.}~\bibnamefont {Bin}}, \bibinfo {author} {\bibfnamefont {M.}~\bibnamefont {Filianina}}, \bibinfo {author} {\bibfnamefont {M.}~\bibnamefont {Dargasz}}, \bibinfo {author} {\bibfnamefont {N.~D.}\ \bibnamefont {Anthuparambil}}, \bibinfo {author} {\bibfnamefont {J.}~\bibnamefont {M{\"o}ller}}, \bibinfo {author} {\bibfnamefont {A.}~\bibnamefont {Zozulya}}, \bibinfo {author} {\bibfnamefont {I.}~\bibnamefont {Andronis}}, \bibinfo {author} {\bibfnamefont {S.}~\bibnamefont {Timmermann}}, \bibinfo {author} {\bibfnamefont {S.}~\bibnamefont {Berkowicz}}, \bibinfo {author} {\bibfnamefont {S.}~\bibnamefont {Retzbach}}, \bibinfo {author} {\bibfnamefont {M.}~\bibnamefont {Reiser}}, \bibinfo {author} {\bibfnamefont {A.~M.}\ \bibnamefont {Raza}}, \bibinfo {author} {\bibfnamefont {M.}~\bibnamefont {Kowalski}}, \bibinfo {author} {\bibfnamefont {M.~S.}\ \bibnamefont {Akhundzadeh}}, \bibinfo {author} {\bibfnamefont {J.}~\bibnamefont {Schrage}}, \bibinfo {author} {\bibfnamefont {C.~H.}\ \bibnamefont {Woo}}, \bibinfo {author} {\bibfnamefont {M.~D.}\ \bibnamefont {Senft}}, \bibinfo {author} {\bibfnamefont {L.~F.}\ \bibnamefont {Reichart}}, \bibinfo {author} {\bibfnamefont {A.}~\bibnamefont {Leonau}}, \bibinfo {author} {\bibfnamefont {P.~R.}\ \bibnamefont {Prince}}, \bibinfo {author} {\bibfnamefont {W.}~\bibnamefont {Ch{\`e}vremont}}, \bibinfo {author} {\bibfnamefont {T.}~\bibnamefont {Seydel}}, \bibinfo {author} {\bibfnamefont {J.}~\bibnamefont {Hallmann}}, \bibinfo {author} {\bibfnamefont {A.}~\bibnamefont {Rodriguez-Fernandez}}, \bibinfo {author} {\bibfnamefont {J.-E.}\ \bibnamefont {Pudell}}, \bibinfo {author} {\bibfnamefont {F.}~\bibnamefont {Brausse}}, \bibinfo {author} {\bibfnamefont {U.}~\bibnamefont {Boesenberg}}, \bibinfo {author} {\bibfnamefont {J.}~\bibnamefont {Wrigley}}, \bibinfo {author} {\bibfnamefont {M.}~\bibnamefont {Youssef}}, \bibinfo {author} {\bibfnamefont {W.}~\bibnamefont {Lu}}, \bibinfo
  {author} {\bibfnamefont {W.}~\bibnamefont {Jo}}, \bibinfo {author} {\bibfnamefont {R.}~\bibnamefont {Shayduk}}, \bibinfo {author} {\bibfnamefont {T.}~\bibnamefont {Guest}}, \bibinfo {author} {\bibfnamefont {A.}~\bibnamefont {Madsen}}, \bibinfo {author} {\bibfnamefont {F.}~\bibnamefont {Lehmk{\"u}hler}}, \bibinfo {author} {\bibfnamefont {M.}~\bibnamefont {Paulus}}, \bibinfo {author} {\bibfnamefont {F.}~\bibnamefont {Zhang}}, \bibinfo {author} {\bibfnamefont {F.}~\bibnamefont {Schreiber}}, \bibinfo {author} {\bibfnamefont {C.}~\bibnamefont {Gutt}},\ and\ \bibinfo {author} {\bibfnamefont {F.}~\bibnamefont {Perakis}},\ }\bibfield  {title} {\enquote {\bibinfo {title} {{Coherent X-rays reveal anomalous molecular diffusion and cage effects in crowded protein solutions}},}\ }\href {https://doi.org/10.1038/s41467-025-66972-6} {\bibfield  {journal} {\bibinfo  {journal} {Nat. Comm.}\ }\textbf {\bibinfo {volume} {16}},\ \bibinfo {pages} {10814} (\bibinfo {year} {2025})}\BibitemShut {NoStop}%
\bibitem [{\citenamefont {Shin}\ \emph {et~al.}(2010)\citenamefont {Shin}, \citenamefont {Kim}, \citenamefont {Talkner},\ and\ \citenamefont {Lee}}]{Shin:2010aa}%
  \BibitemOpen
  \bibfield  {author} {\bibinfo {author} {\bibfnamefont {H.~K.}\ \bibnamefont {Shin}}, \bibinfo {author} {\bibfnamefont {C.}~\bibnamefont {Kim}}, \bibinfo {author} {\bibfnamefont {P.}~\bibnamefont {Talkner}},\ and\ \bibinfo {author} {\bibfnamefont {E.~K.}\ \bibnamefont {Lee}},\ }\bibfield  {title} {\enquote {\bibinfo {title} {{Brownian motion from molecular dynamics}},}\ }\href {https://doi.org/10.1016/j.chemphys.2010.05.019} {\bibfield  {journal} {\bibinfo  {journal} {Chem. Phys.}\ }\textbf {\bibinfo {volume} {375}},\ \bibinfo {pages} {316--326} (\bibinfo {year} {2010})}\BibitemShut {NoStop}%
\bibitem [{\citenamefont {Balucani}\ and\ \citenamefont {Zoppi}(1994)}]{BalucaniBook}%
  \BibitemOpen
  \bibfield  {author} {\bibinfo {author} {\bibfnamefont {U.}~\bibnamefont {Balucani}}\ and\ \bibinfo {author} {\bibfnamefont {M.}~\bibnamefont {Zoppi}},\ }\href@noop {} {\emph {\bibinfo {title} {{Dynamics of the Liquid Phase}}}}\ (\bibinfo  {publisher} {Clarendon Press},\ \bibinfo {address} {Oxford},\ \bibinfo {year} {1994})\BibitemShut {NoStop}%
\bibitem [{\citenamefont {Minh}, \citenamefont {Varghese},\ and\ \citenamefont {Rotenberg}(2026)}]{10.1063/5.0323816}%
  \BibitemOpen
  \bibfield  {author} {\bibinfo {author} {\bibfnamefont {T.~H.~N.}\ \bibnamefont {Minh}}, \bibinfo {author} {\bibfnamefont {S.}~\bibnamefont {Varghese}},\ and\ \bibinfo {author} {\bibfnamefont {B.}~\bibnamefont {Rotenberg}},\ }\bibfield  {title} {\enquote {\bibinfo {title} {{Coupled concentration-charge dynamics in 1:1 electrolytes with unequal diffusion coefficients: Local transient response and fluctuations}},}\ }\href {https://doi.org/10.1063/5.0323816} {\bibfield  {journal} {\bibinfo  {journal} {J. Chem. Phys.}\ }\textbf {\bibinfo {volume} {164}},\ \bibinfo {pages} {194109} (\bibinfo {year} {2026})}\BibitemShut {NoStop}%
\bibitem [{\citenamefont {Howse}\ \emph {et~al.}(2007)\citenamefont {Howse}, \citenamefont {Jones}, \citenamefont {Ryan}, \citenamefont {Gough}, \citenamefont {Vafabakhsh},\ and\ \citenamefont {Golestanian}}]{Howse:2007aa}%
  \BibitemOpen
  \bibfield  {author} {\bibinfo {author} {\bibfnamefont {J.~R.}\ \bibnamefont {Howse}}, \bibinfo {author} {\bibfnamefont {R.~A.~L.}\ \bibnamefont {Jones}}, \bibinfo {author} {\bibfnamefont {A.~J.}\ \bibnamefont {Ryan}}, \bibinfo {author} {\bibfnamefont {T.}~\bibnamefont {Gough}}, \bibinfo {author} {\bibfnamefont {R.}~\bibnamefont {Vafabakhsh}},\ and\ \bibinfo {author} {\bibfnamefont {R.}~\bibnamefont {Golestanian}},\ }\bibfield  {title} {\enquote {\bibinfo {title} {{Self-motile colloidal particles: From directed propulsion to random walk}},}\ }\href {https://doi.org/10.1103/physrevlett.99.048102} {\bibfield  {journal} {\bibinfo  {journal} {Phys.\ Rev.\ Lett.}\ }\textbf {\bibinfo {volume} {99}},\ \bibinfo {pages} {048102} (\bibinfo {year} {2007})}\BibitemShut {NoStop}%
\end{thebibliography}%

\end{document}